\documentclass[a4paper,UKenglish,cleveref, autoref, thm-restate]{lipics-v2021}
\pdfoutput=1 %uncomment to ensure pdflatex processing (mandatatory e.g. to submit to arXiv)
\usepackage{cedilleverbatim}
\usepackage{proof}
\mathchardef\mhyph="2D % Define a "math hyphen"
\DeclareUnicodeCharacter{2245}{\ensuremath{\cong}}
\newcommand{\abs}[4]{{#1}\, #2\!: \! #3.\, #4}
\newcommand{\absu}[3]{{#1}\, #2.\, #3}

\title{Simulating Large Eliminations in Cedille} %TODO Please add

\author{%
  Christopher Jenkins}{%
  The University of Iowa, U.S.A.\and \url{https://homepage.divms.uiowa.edu/{\textasciitilde}cwjnkins/} }{%
  christopher-jenkins@uiowa.edu}{https://orcid.org/0000-0002-5434-5018}{}

\author{%
  Andrew Marmaduke}{%
  The University of Iowa, U.S.A.\and \url{https://homepage.divms.uiowa.edu/{\textasciitilde}marmaduke/} }{%
  andrew-marmaduke@uiowa.edu}{}{}

\author{%
  Aaron Stump}{%
  The University of Iowa, U.S.A.\and \url{https://homepage.divms.uiowa.edu/{\textasciitilde}astump/} }{%
  aaron-stump@uiowa.edu}{}{}

\authorrunning{C.\ Jenkins and A.\ Marmaduke and A.\ Stump}
\hideLIPIcs
\category{} %optional, e.g. invited paper

\relatedversion{} %optional, e.g. full version hosted on arXiv, HAL, or other respository/website
\supplement{\url{https://github.com/cedille/cedille-developments/}}
\nolinenumbers %uncomment to disable line numbering

\begin{document}

\maketitle

\begin{abstract}
  Large eliminations provide an expressive mechanism for arity- and type-generic
  programming.
  However, as large eliminations are closely tied to a type theory's primitive
  notion of inductive type, this expressivity is not expected within polymorphic
  lambda calculi in which datatypes are encoded using impredicative quantification.
  We report progress on simulating large eliminations for datatype encodings in
  one such type theory, the \emph{calculus of dependent lambda eliminations}
  (CDLE).
  Specifically, we show that the expected computation rules for large
  eliminations, expressed using a derived type of extensional equality of
  types, can be proven \emph{within} CDLE.
  We present several case studies, demonstrating the adequacy of this
  simulation for a variety of generic programming tasks, and a generic
  formulation of the simulation allowing its use for any datatype.
  All results have been mechanically checked by Cedille, an implementation of CDLE.
\end{abstract}

\section{Introduction}
\label{sec:intro}
In dependently typed languages, large eliminations allow programmers
to define types by induction over datatypes --- that is, as an elimination of
a datatype into the large universe of types.
For type theory semanticists, large eliminations rule out two-element
models of types by providing a principle of proof discrimination (e.g., \(0 \neq
1\))\cite{Smi88_Independence-of-Peanos-4th-Axiom-from-MLTT,
  Smi84_An-Interpretation-of-MLTT-in-a-Type-Free-Theory}.
For programmers, they give an expressive mechanism for arity- and type-generic
programming with universe
constructions~\cite{WC10_Generic-Programming-with-Dependent-Types}.
As an example, the type \(\mathit{Nary}\ n\) of \(n\)-ary functions (where \(n\)
is a natural number) over some type \(T\) can be defined as \(T\) when \(n = 0\)
and \(T \rightarrow \mathit{Nary}\ n'\) when \(n = 1 + n'\).

Large eliminations are closely tied to a type theory's primitive notion of
inductive type, and thus this expressivity is not expected within polymorphic
pure typed lambda calculi in which datatypes are impredicatively encoded.
The \emph{calculus of dependent lambda eliminations} (CDLE)~\cite{Stu17_CDLE,
  SJ20_Cedille-Syntax-Semantics} is one such theory that seeks to overcome
historical difficulties of impredicative encodings, such as the lack of induction
principles for datatypes~\cite{FS18_Generic-Derivation-of-Induction-in-Cedille}.

\subparagraph*{Contributions}
In this paper, we report progress on overcoming another difficulty of
impredicative encodings: the lack of large eliminations.
We show that the expected definitional equalities of a
large elimination can be simulated using a \emph{derived type of
extensional equality} for types (where the extent of a type is the set of terms
it classifies).
In particular, we:
\begin{itemize}
\item describe a method for simulating large eliminations in CDLE
  (Section~\ref{sec:nary}), identifying the features of the theory that enable
  the development (Section~\ref{sec:cdle-background});
  
\item formulate the method \emph{generically} (meaning \emph{parametrically})
  for all datatype signatures, using the framework of Firsov et
  al.~\cite{FBS18_Efficient-Mendler-Style-Lambda-Encodings}
  (Section~\ref{sec:generic-large-elim});
  
\item demonstrate the adequacy of this simulation by applying it to several
  generic programming tasks: \(n\)-ary functions, a closed universe of
  datatypes, a decision procedure for the inhabitation of simple types, and an
  arity-generic map operation (Sections~\ref{sec:nary}
  and~\ref{sec:case-studies}).
\end{itemize}
All results have been mechanically checked by Cedille, an implementation of
CDLE, and are available at the code repository associated with this
paper.\footnote{\url{https://github.com/cedille/cedille-developments/tree/master/large-elim-sim}}

\subparagraph*{Outline}
The remainder of this paper is structured as follows.
Section~\ref{sec:cdle-background} reviews background material on CDLE, focusing
on the primitives which enable the simulation and the derived extensional type
equality.
In Section~\ref{sec:nary}, we carefully explain the recipe for simulating large
eliminations using as an example the type \(n\)-ary functions over a given type.
Section~\ref{sec:case-studies} shows three case studies, presented as evidence
of the effectiveness of the simulation in tackling generic programming tasks.
The recipe for concrete examples is then turned into a generic (that is,
parametric) derivation of simulated large eliminations in
Section~\ref{sec:generic-large-elim}.
Finally, Section~\ref{sec:related} discusses related work and
Section~\ref{sec:conclusion} concludes with a discussion of future work.

\section{Background on CDLE}
\label{sec:cdle-background}
In this section, we review CDLE, the kernel theory of Cedille.
CDLE extends the impredicative extrinsically typed \emph{calculus of
  constructions} (CC), overcoming historical difficulties of impredicative
encodings (e.g., underivability of
induction~\cite{Geu01_Induction-is-not-derivable-in-2O}) by adding three new 
type constructs: equality of untyped terms; the dependent intersections of
Kopylov~\cite{Kop03}; and the implicit products of Miquel~\cite{Mi01_ICC}.
The pure term language of CDLE is untyped lambda calculus,
but to make type checking algorithmic terms are presented with typing
annotations.
Definitional equality of terms \(t_1\) and \(t_2\) is \(\beta\eta\)-equivalence
modulo erasure of annotations, denoted \(|t_1| =_{\beta\eta} |t_2|\).

The typing and erasure rules for the fragment of CDLE used in this paper are shown
in Figure~\ref{fig:cdle-primitives} and described in
Section~\ref{sec:primitives} (see also Stump and
Jenkins~\cite{SJ20_Cedille-Syntax-Semantics}); the derived constructs we use are
presented axiomatically in Section~\ref{sec:derived-constructs}.
We assume the reader is familiar with the type constructs inherited from CC:
abstraction over types in terms is written \(\absu{\Lambda}{X}{t}\) (erasing to
\(|t|\)), application of terms to types (polymorphic type instantiation) is
written \(t \cdot T\) (erasing to \(|t|\)), and application of type constructors
to type constructors is written \(T_1 \cdot T_2\).
In code listings, we sometimes omit type arguments to terms when Cedille can
infer them.

\subsection{Primitives}
\label{sec:primitives}

Below, we only discuss implicit products and the equality type.
Though dependent intersections play a critical role in the derivation of
induction for datatype encodings, they are otherwise not explicitly used in the
coming developments.

\begin{figure}
  \centering
  \[
    \begin{array}{c}
    \begin{array}{cc}
      \infer{\Gamma\vdash \absu{\Lambda}{x}{t} : \abs{\forall}{x}{T_1}{T_2}}
      {
        \Gamma,x:T_1\vdash t : T_2
        \quad x\not\in\textit{FV}(|t|)
      }
      &
        \infer{
         \Gamma\vdash t\ \mhyph t' : [t'/x]T_2
        }{
        \Gamma\vdash t : \abs{\forall}{x}{T_1}{T_2}
        \quad \Gamma\vdash t' : T_1
        }
      \\ \\
      \infer{\Gamma\vdash \beta : \{t_1 \simeq t_2\}}
      {
        |t_1| =_{\beta\eta} |t_2|
      }
      &
        \infer{
        \Gamma \vdash \delta\ \mhyph\ t : T
        }{
        \Gamma \vdash t :
        \{\absu{\lambda}{x}{\absu{\lambda}{y}{x}} \simeq \absu{\lambda}{x}{\absu{\lambda}{y}{y}}\}
         }
    \end{array}
      \\ \\
      \infer{
      \Gamma\vdash \varphi\ t\ \mhyph\ t'\ \{t''\} : T
      }{
      \Gamma\vdash t : \{t' \simeq t''\}
      \quad \Gamma\vdash t' : T
      \quad \textit{FV}(t'') \subseteq \textit{dom}(\Gamma)
      }
      \\ \\
      \begin{array}{lllllll}
        |\absu{\Lambda}{x}{t}| & = & |t| &\
        & |t\ \mhyph t'| & = & |t|
        \\ |\beta| & = & \absu{\lambda}{x}{x} &\ 
        & |\varphi\ t\ \mhyph\ t'\ \{t''\}| & = & |t''|
        \\ |\delta\ \mhyph\ t| & = & \absu{\lambda}{x}{x}
      \end{array}
    \end{array}
  \]
  \caption{Typing and erasure for a fragment of CDLE}
  \label{fig:cdle-primitives}
\end{figure}

\subparagraph*{The implicit product type \(\abs{\forall}{x}{T_1}{T_2}\)} of
Miquel~\cite{Mi01_ICC} is the type for functions which accept an erased
(computationally irrelevant) input of type \(T_1\) and produce a result of type
\(T_2\).
Implicit products are introduced with \(\absu{\Lambda}{x}{t}\), and the type
inference rule is the same as for ordinary function abstractions except for the side
condition that \(x\) does not occur free in the erasure of the body \(t\).
Thus, the argument plays no computational role in the function and exists
solely for the purposes of typing.
The erasure of the introduction form is \(|t|\).
For application, if \(t\) has type \(\abs{\forall}{x}{T_1}{T_2}\) and \(t'\) has
type \(T_1\), then \(t\ \mhyph t'\) has type \([t'/x]T_2\) and erases to
\(|t|\).
When \(x\) is not free in \(T_2\), we write \(T_1 \Rightarrow T_2\),
similar to writing \(T_1 \to T_2\) for \(\abs{\Pi}{x}{T_1}{T_2}\). 

\begin{note*}
  The notion of computational irrelevance here is not that of a
  different sort of classifier for types (e.g. \(\mathit{Prop}\) in Coq,
  c.f.~\cite{Coq12}) separating terms by whether they can be used for computation.
  Instead, it is similar to \emph{quantitative type
    theory}~\cite{Atk18_Quantitative-Type-Theory}: relevance and irrelevance are
  properties of binders, indicating how functions may use arguments.
\end{note*}

\subparagraph*{The equality type \(\{t_1 \simeq t_2\}\)} is the type of
proofs that \(t_1\) is propositionally equal to \(t_2\).
The introduction form \(\beta\) proves reflexive equations between
\(\beta\eta\)-equivalence classes of terms: it can be checked against the type
\(\{t_1 \simeq t_2\}\) if \(|t_1| =_{\beta\eta} |t_2|\).
Note that this means equality is over \emph{untyped} (post-erasure) terms.
There is also a standard elimination form (substitution), but it is not used
explicitly in the presentation of our results so we omit its inference rule.

Equality types also come with two additional axioms: a strong form of the
direct computation rule of NuPRL (c.f.\ Allen et
al.~\cite{ABCEKLM06_Innovations-in-Computational-TT-using-NuPRL}, Section~2.2)
given by \(\varphi\), and a principle of proof discrimination given by \(\delta\).
The inference rule for an expression of the form \(\varphi\ t\ \mhyph\ t'\
\{t''\}\) says that the entire expression can be checked against type \(T\) if
\(t'\) can be, if there are no undeclared free variables in \(t''\) (so, \(t''\)
is a well-scoped but otherwise untyped term), and if \(t\) proves that \(t'\)
and \(t''\) are equal.
The crucial feature of \(\varphi\) is its erasure: the expression erases
to \(|t''|\), effectively enabling us to cast \(t''\) to the type of \(t'\).
Though \(\varphi\) does not appear explicitly in the developments to come, it
plays a central role by enabling the derivation of extensional type equality.

An expression of the form \(\delta\ \mhyph\ t\) may be checked against any type
if \(t\) synthesizes a type convertible with a particular false equation,
\(\{\absu{\lambda}{x}{\absu{\lambda}{y}{x}} \simeq
\absu{\lambda}{x}{\absu{\lambda}{y}{y}}\}\).
To broaden applicability of \(\delta\), the Cedille tool implements the
\emph{B\"ohm-out} semi-decision procedure \cite{BDPR79_Bohm-Algorithm} for
discriminating between \(\beta\eta\)-inequivalent terms.
By enabling proofs that datatype constructors are disjoint, \(\delta\) plays a
vital role in our simulation of large eliminations.

\subsection{Derived Constructs}
\label{sec:derived-constructs}

\begin{figure}[h]
  \centering
  \[
    \begin{array}{c}
      \begin{array}{cc}
        \infer{
        \Gamma \vdash \mathit{intrCast} \cdot S \cdot T\ \mhyph t_1\ \mhyph t_2
        : \mathit{Cast} \cdot S \cdot T
        }{
        \Gamma \vdash t_1 : S \to T
        \quad \Gamma \vdash t_2 : \abs{\Pi}{x}{S}{\{t_1\ x \simeq x\}}
        }
        &
          \infer{
          \Gamma \vdash \mathit{cast} \cdot S \cdot T\ \mhyph t : S \to T
          }{
          \Gamma \vdash t : \mathit{Cast} \cdot S \cdot T
          }
      \end{array}
      \\ \\
      \begin{array}{lllllll}
        |\mathit{intrCast} \cdot S \cdot T\ \mhyph t_1\ \mhyph t_2| & = & \absu{\lambda}{x}{x} &\ 
        & |\mathit{cast} \cdot S \cdot T\ \mhyph t| & = & \absu{\lambda}{x}{x}
      \end{array}
    \end{array}
  \]
  \caption{Type inclusions}
  \label{fig:cast}
\end{figure}

\paragraph*{Type inclusions}
The \(\varphi\) axiom of equality allows us to define a type
constructor \(\mathit{Cast}\) which internalizes the notion that the set of all
elements of some type \(S\) is contained within the set of all elements of type
\(T\) (note that Curry-style typing makes this relation nontrivial).
We describe its axiomatic summary presented in Figure~\ref{fig:cast}; for the
full derivation, see Jenkins and
Stump~\cite{JS20_Monotone-Recursive-Types-in-Cedille} (also Diehl et
al.~\cite{DFS18_Generic-Zero-Cost-Reuse}).

The introduction form \(\mathit{intrCast}\) takes two erased term arguments, a
function \(t_1 : S \to T\), and a proof that \(t_1\) behaves extensionally as
the identity function on its domain.
The elimination form \(\mathit{cast}\) takes evidence that a type \(S\) is
included into \(T\) and produces a function between the same.
The crucial property of \(\mathit{cast}\) is its erasure: \(|\mathit{cast}\
\mhyph t| = \absu{\lambda}{x}{x}\).
Thus, \(\mathit{Cast} \cdot S \cdot T\) may also be considered the type of
\emph{zero-cost} type coercions from \(S\) to \(T\) --- zero cost because the
type coercion is performed in a constant number of \(\beta\)-reduction steps.

\begin{note*}
  When inspecting the introduction and elimination forms, it may seem that
  \(\mathit{Cast}\) provides a form of function extensionality restricted to
  identity functions.
  This is not the case, however, as it is possible to choose \(S\), \(T\), and
  \(t_1\) such that \(t_1\) is provably extensionally equal to the identity
  function on its domain, and \emph{at the same time} refute \(\{ t_1 \simeq
  \absu{\lambda}{x}{x}\}\) using \(\delta\).
  Instead, taken together these rules should be read as saying that if \(t_1\) is
  extensionally identity on its domain, then that justifies the
  \emph{assignment of type \(S \to T\) to \(\absu{\lambda}{x}{x}\)}.
\end{note*}

\begin{figure}[h]
  \centering
  \[
    \begin{array}{c}
        \infer{
         \Gamma \vdash \mathit{intrTpEq} \cdot S \cdot T\ \mhyph t_1\ \mhyph t_2
         : \mathit{TpEq} \cdot S \cdot T
        }{
          \Gamma \vdash t_1 : \mathit{Cast} \cdot S \cdot T
          \quad \Gamma \vdash t_2 : \mathit{Cast} \cdot T \cdot S
        }
      \\ \\
      \begin{array}{cc}
        \infer{
         \Gamma \vdash tpEq1 \cdot S \cdot T\ \mhyph t : S \to T
        }{
         \Gamma \vdash t : \mathit{TpEq} \cdot S \cdot T
        }
        &
          \infer{
           \Gamma \vdash tpEq2 \cdot S \cdot T\ \mhyph t : T \to S
          }{
           \Gamma \vdash t : \mathit{TpEq} \cdot S \cdot T
          }
      \end{array}
      \\ \\
      \begin{array}{lllllll}
        |\mathit{intrTpEq} \cdot S \cdot T\ \mhyph t_1\ \mhyph t_2| & = & \absu{\lambda}{x}{x} &\ 
        & |\mathit{tpEq1} \cdot S \cdot T\ \mhyph t| & = & \absu{\lambda}{x}{x}
        \\ |\mathit{tpEq2} \cdot S \cdot T\ \mhyph t| & = & \absu{\lambda}{x}{x}
      \end{array}          
    \end{array}
  \]
  \caption{Extensional type equality}
  \label{fig:tpeq}
\end{figure}

\paragraph*{Type equality}
The extensional notion of type equality used to simulate large eliminations,
\(\mathit{TpEq}\), is merely the existence of a two-way type inclusion.
Strictly speaking, a term of type \(\mathit{TpEq} \cdot S \cdot T\) is evidence
that \(\absu{\lambda}{x}{x}\) may be assigned the intersection type \((S \to T)
\cap (T \to S)\).
However, a good intuition for understanding the introduction and elimination
rules is to think of this as the types of pairs of casts \(t_1\) and \(t_2\)
between \(S\) and \(T\), with the elimination forms being compositions of the
appropriate projection function with \(\mathit{cast}\) (the projections would
not appear in the erasures of the elimination forms, as the argument to
\(\mathit{cast}\) is erased).

\begin{remark}
  \label{rem:tpeq-subst}
  Though we call \(\mathit{TpEq}\) extensional type \emph{equality}, within CDLE
  it is only an isomorphism of types.
  To be considered a true notion of equality, \(\mathit{TpEq}\) would need a
  substitution principle.
  The type constructors for dependent function types (both implicit and
  explicit) can be proven to permit substitution if the domain and codomain
  parts do, as does quantification over types.
  However, the presence of higher-order type constructors means that not all
  closed types allow substitution of types related by \(\mathit{TpEq}\)
  (consider \(\abs{\forall}{X}{\star \to \star}{X \cdot S}\), where one has
  \(S\) and \(T\) such that \(\mathit{TpEq} \cdot S \cdot T\)).
  Nonetheless, the case studies presented in Sections~\ref{sec:nary}
  and~\ref{sec:case-studies} show that despite this limitation, our simulation
  of large eliminations using \(\mathit{TpEq}\) is adequate for dealing with
  common generic-programming tasks.
\end{remark}

\begin{note*}
  In the developments in subsequent sections, \(\mathit{refl}\),
  \(\mathit{sym}\), and \(\mathit{trans}\) refer to the three proofs that
  together show \(\mathit{TpEq}\) is an equivalence relation.
  We have omitted their definitions; their types are as expected.
\end{note*}

\section{\(n\)-ary Functions}
\label{sec:nary}

\begin{figure}
  \centering
  \small
  \begin{subfigure}{0.41\linewidth}
    \caption{As a large elimination}
    \label{fig:nary-le}
\begin{verbatim}
Nary : Nat ➔ ★
Nary zero = T
Nary (succ n) = T ➔ Nary n

\end{verbatim}
  \end{subfigure}%
  \begin{subfigure}{0.59\linewidth}
    \caption{As a GADT}
    \label{fig:nary-rel}
\begin{verbatim}
data NaryR : Nat ➔ ★ ➔ ★
= naryRZ : NaryR zero T
| naryRS : ∀ n: Nat. ∀ Ih: ★.
           NaryR n ·Ih ➔ NaryR (succ n) ·(T ➔ Ih)
\end{verbatim}
  \end{subfigure}
  \caption{\(n\)-ary functions over \(T\)}
  \label{fig:cdle-delta-eq}
\end{figure}

In this section, we use a concrete example to detail the method of simulating
large eliminations.
Figure~\ref{fig:nary-le} shows the definition of \(\mathit{Nary}\), the family of
\(n\)-ary function types over some type \(T\), as a large elimination of natural
numbers.
Our simulation of this begins by approximating this inductive definition of
a \emph{function} with an inductive \emph{relation} between \(\mathit{Nat}\) and
types, given as the generalized algebraic
datatype~\cite{XCC03_Guarded-Recursive-Datatype-Constructors} (GADT)
\(\mathit{NaryR}\) in Figure~\ref{fig:nary-rel}.

This approximation is inadequate: we lack a canonical name for the type
\(\mathit{Nary}\ n\) because \(n\) does not \emph{a priori} determine the type
argument of \(\mathit{NaryR}\ n\).
Indeed, without a form of proof discrimination we would not even be able to
deduce that if a given type \(N\) satisfies \(\mathit{NaryR}\ \mathit{zero}\),
then \(N\) must be \(T\).
Proceeding by induction, in the \(\mathit{naryRS}\) case the goal is to show
that \(T\) is the same as \(T \to \mathit{Ih}\) for some (arbitrary but fixed)
\(\mathit{Ih} : \star\).
We would need to derive a contradiction from the absurd equation that
\(\{\mathit{zero} \simeq \mathit{succ}\ n\}\) for some \(n\).
Fortunately, proof discrimination \emph{is} available in CDLE in the form of
\(\delta\), so we are able to define functions such as \(\mathit{extr0}\) in
Figure~\ref{fig:delta1} which require this form of reasoning.

\begin{figure}[h]
  \centering
  \small
\begin{verbatim}
extr0' : ∀ x: Nat. { x ≃ zero } ➾ ∀ N: ★. NaryR x ·N ➔ N ➔ T
extr0' -zero -eqx ·T naryRZ x = x
extr0' -(succ n) -eqx ·(T → X) (naryRS n ·X r) x = δ - eqX

extr0 = extr0' -zero -β
\end{verbatim}
  \caption{Extracting a \(0\)-ary term}
  \label{fig:delta1}
\end{figure}

\begin{note*}
  In code listings such as Figure~\ref{fig:delta1}, we present recursive Cedille
  functions using the syntax of (dependent) pattern matching in order to aid readability.
  This syntax is not currently supported by the Cedille tool.
  For functions computing terms, the Cedille code in this paper's repository uses
  the datatype system described by Jenkins et
  al.~\cite{JMS20_Elaborating-Inductive-Definitions-and-COV-Induction-Cedille}.
  For functions computing types, the repository code uses the simulation to be
  described next.
\end{note*}

\paragraph*{Sketch of the Idea}
Our task is to show that \(\mathit{NaryR}\) defines a \emph{functional}
relation, i.e., for all \(n : \mathit{Nat}\) there exists a unique type
\(\mathit{Nary}\ n\) such that \(\mathit{NaryR}\ n \cdot (\mathit{Nary}\ n)\) is
inhabited.
The candidate definition for this type family is:
\begin{verbatim}
Nary n = ∀ X: ★. NaryR n ·X ➾ X
\end{verbatim}
For all \(n\), read \(\mathit{Nary}\ n\) as the type of
terms contained in the intersection of the family of types \(X\) such that
\(\mathit{NaryR}\ n \cdot X\) is inhabited.
For example, every term \(t\) of type \(\mathit{Nary}\ \mathit{zero}\) has type
\(T\) also, since \(T\) is in this family (specifically, we have that \(t \cdot
T\ \mhyph \mathit{naryRZ}\) has type \(T\) and erases to \(|t|\)).
Similarly, every term of type \(T\) has type \(\mathit{Nary}\ \mathit{zero}\) by
induction on the assumed proof of \(\mathit{NaryR}\ \mathit{zero} \cdot X\) for
arbitrary \(X\), invoking \(\delta\) in the \(\mathit{naryRS}\) case (see 
Section~\ref{sec:computation-laws-coercions} for how certain properties may be
proved using induction on erased arguments).

However, at the moment we are stuck when attempting to prove \(\mathit{NaryR}\
\mathit{zero} \cdot (\mathit{Nary}\ \mathit{zero})\).
Though we see from the preceding discussion that \(T\) and \(\mathit{Nary}\
\mathit{zero}\) are \emph{extensionally} equal types (they classify the same
terms), \(\mathit{naryRZ}\) requires that they be \emph{definitionally} equal!
We therefore must modify the definition of \(\mathit{NaryR}\) so that it defines
a relation that is functional with respect to extensional type equality.
This is shown in Figure~\ref{fig:nary-frel}, with both constructors now
quantifying over an additional type argument \(X\) together with evidence that
it is extensionally equal to the type of interest.

\begin{figure}[h]
  \centering
  \small
\begin{verbatim}
data NaryR : Nat ➔ ★ ➔ ★
= naryRZ : ∀ X: ★. TpEq ·X ·T ➾ NaryR zero ·X
| naryRS : ∀ Ih: ★. ∀ n: Nat. NaryR n ·Ih ➔
           ∀ X: ★. TpEq ·X ·(T ➔ Ih) ➾ NaryR (succ n) ·X
\end{verbatim}
  \caption{\(\mathit{NaryR}\) as a relation that is functional with respect to \(\mathit{TpEq}\)}
  \label{fig:nary-frel}
\end{figure}

\subsection{Proof that \(\mathit{NaryR}\) is a Functional Relation}
\paragraph*{Respectfulness}
We now detail the proof that \(\mathit{NaryR}\) is a functional relation.
Having changed our notion of equality for types to extensional, we must now
prove a third property (in addition to uniqueness and existence): that it
respects extensional type equality, i.e., if \(\mathit{Nary}\) relates \(n\) to
\(T_1\) and \(T_1\) is equal to \(T_2\), then \(\mathit{Nary}\) relates \(n\) to
\(T_2\) also.
This proof is shown as \(\mathit{naryRResp}\) in Figure~\ref{fig:nary-resp}.
Proceeding by case analysis, in both cases we combine the assumed proof that
\(T_1\) and \(T_2\) are equal types with the type equality proof given to the
constructor.

\begin{figure}[h]
  \centering
  \small
\begin{verbatim}
naryRResp : ∀ n: Nat. ∀ T1: ★. NaryR n ·T1 ➔ ∀ T2: ★. TpEq ·T1 ·T2 ➾ NaryR n ·T2
naryRResp -zero ·T1 (naryRZ ·T1 -eqT1) ·T2 -eq =
  naryRZ ·T2 -(trans -(sym -eq) -eqT1)
naryRResp -(succ n) ·T1 (naryRS ·Ih -n r ·T1 -eqT1) ·T2 -eq =
  naryRS ·Ih -n r ·T2 -(trans -(sym -eq) -eqT1)
\end{verbatim}
  \caption{\(\mathit{NaryR}\ n\) respects type equality}
  \label{fig:nary-resp}
\end{figure}

\paragraph*{Uniqueness}
Figure~\ref{fig:nary-unique} shows the proof \(\mathit{naryRUnique}\) that
\(\mathit{Nary}\ n\) uniquely determines a type (up to type equality) for all
\(n\).
To improve readability, this listing omits the two absurd clauses in which the
given \(\mathit{Nary}\) proofs differ in their construction; in the code
repository, these two cases are handled with \(\delta\).
For the \(\mathit{naryRS}\) case, the inductive hypothesis gives us that
\(\mathit{Ih}_1\) and \(\mathit{Ih}_2\) are equal types.
We combine the lemma \(\mathit{arrowTpEqCod}\) (definition omitted) that the
arrow type constructor respects type equality in its codomain with the proofs
given to the constructors that \(T \to \mathit{Ih}_1\) is equal to \(T_1\) and
\(T \to \mathit{Ih}_2\) is equal to \(T_2\), concluding the proof.

\begin{figure}[h]
  \centering
  \small
\begin{verbatim}
arrowTpEqCod : ∀ D: ★. ∀ C1: ★. ∀ C2: ★. TpEq ·C1 ·C2 ➾ TpEq ·(D ➔ C1) ·(D ➔ C2)

naryRUnique : ∀ n: Nat. ∀ T1: ★. NaryR n ·T1 ➔ ∀ T2: ★. NaryR n ·T2 ➔ TpEq ·T1 ·T2
naryRUnique -zero ·T1 (naryRZ ·T1 -eqT1) ·T2 (naryRZ ·T2 -eqT2) =
  trans -eqT2 -(sym -eqT1)
naryRUnique -(succ n) ·T1 (naryRS ·Ih1 -n r1 ·T1 -eqT1) ·T2 (naryRS ·Ih2 -n r2 ·T2 -eqT2) =
  trans -eqT1 -(arrowTpEqCod -(naryRUnique -n r1 r2)) -(sym -eqT2)
\end{verbatim}
  \caption{\(\mathit{NaryR}\ n\) determines at most one type}
    \label{fig:nary-unique}
\end{figure}

\paragraph*{Existence}
Compared to the first two properties, the proof of existence,
\(\mathit{naryREx}\) in Figure~\ref{fig:nary-existence}, is more involved.
We take a top-down approach for its explanation to first impart the main idea.
Proceeding by induction over the natural number argument, the proof relies on
two lemmas: \(\mathit{naryZ}\), which proves that \(\mathit{NaryR}\) relates \(\mathit{zero}\) to
\(\mathit{Nary}\ \mathit{zero}\), and \(\mathit{naryS}\), which proves that
\(\mathit{succ}\ n\) and \(\mathit{Nary}\ (\mathit{succ}\ n)\) are related if
\(n\) and \(\mathit{Nary}\ n\) are.
Put another way, to prove existence we need to specialize the
\(\mathit{naryRZ}\) and \(\mathit{naryRS}\) constructors to the corresponding
members of the \(\mathit{Nary}\) family.

\begin{figure}[h]
  \centering
  \small
\begin{verbatim}
naryREx : Π n: Nat. NaryR n ·(Nary n)
naryREx zero = naryZ
naryREx (succ n) = naryS n (naryREx n)
\end{verbatim}
  \caption{\(\mathit{NaryR}\) relates \(n\) and \(\mathit{Nary}\ n\) for all
    \(n\)}
  \label{fig:nary-existence}
\end{figure}

% \begin{figure}
%   \centering
%   \small
% \begin{verbatim}
% naryRZ' : NaryR zero ·T
% naryRZ' = naryRZ -(refl ·T)

% naryZEq : TpEq ·(Nary zero) ·T
% naryRZEq =
%   intrTpEq -(intrCast -(λ x. x -naryRZ') -(λ _. β))
%     -(intrCast -(λ a. Λ X. Λ rz. tpEq1 -(naryRUnique -zero naryRZ' rz) a) -(λ _. β))

% naryZ : NaryR zero ·(Nary zero)
% naryZ = naryRResp -zero naryRZ' -(sym -naryZEq)
% \end{verbatim}
%   \caption{\(\mathit{Nary}\): \(\mathit{zero}\) case}
%   \label{fig:nary-zero}
% \end{figure}

The proofs of \(\mathit{naryZ}\) and \(\mathit{naryS}\) follow a similar
three-part structure, so for the sake of brevity we detail the proof of the
latter only (Figure~\ref{fig:nary-succ}).
First, with \(\mathit{naryRS'}\) we specialize the constructor
\(\mathit{naryRS}\) to the reflexive type equality.
Then, with \(\mathit{narySEq}\) we prove that the computation rule for
\(\mathit{Nary}\ (\mathit{succ}\ n)\) (c.f.\ Figure~\ref{fig:nary-le}) holds for
all \(n\) such that \(\mathit{NaryR}\ n \cdot (\mathit{Nary}\ n)\) holds.
This is proved as a two-way type inclusion.
\begin{itemize}
\item In the first direction, we assume \(f : \mathit{Nary}\ (\mathit{succ}\
  n)\).
  Since this type is the intersection of the family of types \(S\)
  such \(\mathit{NaryR}\ (\mathit{succ}\ n) \cdot S\) holds, we conclude by
  showing \(T \to \mathit{Nary}\ n\) is in this family.
  By the erasure rules, the first argument to \(\mathit{intrCast}\) erases to
  \(\absu{\lambda}{f}{f}\).
  
\item In the second direction, we assume \(f : T \to \mathit{Nary}\ n\) and an
  arbitrary type \(X\) such that \(\mathit{Nary}\ (\mathit{succ}\ n) \cdot X\)
  holds, and must cast \(f\) to the type \(X\).
  This is done by appealing to uniqueness, as \(\mathit{NaryR}\) relates
  \(\mathit{succ}\ n\) to both \(X\) and \(T \to \mathit{Nary}\ n\).
  Notice that while \(\mathit{rs} : \mathit{Nary}\ (\mathit{succ}\ n) \cdot X\)
  must not occur within the erasure of the body of the \(\Lambda\)-expression
  which binds it, the elimination form \(\mathit{tpEq1}\) takes the type
  equality that \(\mathit{naryRUnique}\) computes from \(\mathit{rs}\) as an
  \emph{erased} argument, ensuring that this condition holds.
\end{itemize}

\noindent Finally, we prove \(\mathit{naryS}\) by appealing to respectfulness.

\begin{figure}
  \centering
  \small
\begin{verbatim}
naryRS' : ∀ Ih: ★. ∀ n: Nat. NaryR n ·Ih ➔ NaryR (succ n) ·(T ➔ Ih)
naryRS' ·Ih -n r = naryRS -n r -(refl ·(T ➔ Ih))

narySEq : ∀ n: Nat. NaryR n ·(Nary n) ➔ TpEq ·(Nary (succ n)) ·(T ➔ Nary n)
narySEq -n rn =
  intrTpEq
    -(intrCast -(λ f. f ·(T ➔ Nary n) -(naryRS' -n rn)) -(λ _. β))
    -(intrCast -(λ f. Λ X. Λ rs.
                 tpEq1 -(naryRUnique -(succ n) (naryRS' -n rn) ·X rs) f)
               -(λ _. β))

naryS : ∀ n: Nat. NaryR n ·(Nary n) ➔ NaryR (succ n) ·(Nary (succ n))
naryS -n rn =
  naryRResp -(succ n) ·(T ➔ Nary n) (naryRS' -n rn) -(sym -(narySEq -n rn))
\end{verbatim}
  \caption{\(\mathit{Nary}\): \(\mathit{succ}\) case}
  \label{fig:nary-succ}
\end{figure}

\subsection{Computation Laws as \emph{Zero-cost} Type Coercions}
\label{sec:computation-laws-coercions}

The proof of existence, \(\mathit{naryREx}\), takes time linear in its argument
\(n\) to compute a proof of \(\mathit{NaryR}\ n \cdot (\mathit{Nary}\ n)\).
Therefore, it would seem that any type coercions using this proof could not be
zero-cost (that is, constant time).
We now show that this issue is neatly dealt with by using the fact that type
equality is \emph{proof-irrelevant}.

Proof irrelevance for a type \(P\) is often understood to mean that any two
proofs of \(P\) are equal.
While type equalities do satisfy this notion of proof irrelevance, in CDLE (and
other theories with usage restrictions on binders), one can formulate an
alternative notion of proof irrelevance: that one can construct a proof of \(P\)
from an \emph{erased} proof of \(P\), i.e., that the type \(P \Rightarrow P\) is
inhabited.

The proof of proof-irrelevance for type equality, and the type equalities
for the computation laws of \(\mathit{Nary}\), are shown in
Figure~\ref{fig:nary-comp}.
In \(\mathit{narySC}\), we invoke the existence proof within an expression given
as an \emph{erased} argument to \(\mathit{tpEqIrrel}\).
Thus, no computation involving \(n\) is performed in the operational semantics
of CDLE when using these type coercions.

\begin{figure}
  \centering
  \small
\begin{verbatim}
tpEqIrrel : ∀ A: ★. ∀ B: ★. TpEq ·A ·B ➾ TpEq ·A ·B
tpEqIrrel ·A ·B -eq =
  intrTpEq -(intrCast -(tpEq1 -eq) -(λ _. β)) -(intrCast -(tpEq2 -eq) -(λ _. β))

naryZC : TpEq ·(Nary zero) ·T
naryZC = naryZEq

narySC : ∀ n: Nat. TpEq ·(Nary (succ n)) ·(T ➔ Nary n)
narySC -n = tpEqIrrel -(narySEq -n (naryREx n))
\end{verbatim}
  \caption{Computation laws of \(\mathit{Nary}\) as type equalities}
  \label{fig:nary-comp}
\end{figure}

We conclude this section with an example: applying an \(n\)-ary function to
\(n\) arguments of type \(T\), given as a length-indexed list (\(\mathit{Vec}\)).
In Figure~\ref{fig:nary-app}, \(\mathit{appN}\) is defined by induction on the
list of arguments.
In the \(\mathit{vcons}\) case, the given natural number is revealed to have the
form \(\mathit{succ}\ n\), so we may coerce the type of \(f : \mathit{Nary}\
(succ\ n)\) to \(T \to \mathit{Nary}\ n\) to apply it to the head of the list,
then recursively call \(\mathit{appN}\) on the tail.

\begin{figure}
  \centering
  \small
\begin{verbatim}
appN : ∀ n: Nat. Nary n ➔ Vec ·T n ➔ T
appN -zero f vnil = tpEq1 -naryZC f
appN -(succ n) f (vcons -n x xs) = appN -n (tpEq1 -(narySC -n) f x) xs
\end{verbatim}
  \caption{Application of an \(n\)-ary function to \(n\) arguments}
  \label{fig:nary-app}
\end{figure}

\section{Generic Programming Case Studies}
\label{sec:case-studies}

In the previous section, we gave a detailed outline of the recipe for
simulating a large elimination, in particular showing explicitly the use of
type coercions.
For the case studies we consider next, all code listings are presented in a
syntax that omits the uses of type coercions to improve readability.
In our implementation, we must explicitly use these coercions as well as several
lemmas showing that CDLE's primitive type constructors respect \(\mathit{TpEq}\)
(except for quantification over higher-kinded type constructors, as mentioned in
Remark~\ref{rem:tpeq-subst}).
As CDLE is a kernel theory (and thus not ergonomic to program in), the purpose
of these examples is to show that this simulation is indeed capable of
expressing common generic programming tasks, and we leave the task of
implementing a high-level surface language for its utilization as future work.
We do, however, remark on any new difficulties that are obscured by this
presentation (such as Remark~\ref{rem:tyvec-eq}).
Full details of all examples of this section can be found in the repository
associated with this paper.

\subsection{A Closed Universe of Strictly Positive Datatypes}
\label{sec:closed-universe}

\begin{figure}[h]
  \centering
  \small
\begin{verbatim}
data Descr : ★
= idD    : Descr
| constD : Descr
| pairD  : Descr ➔ Descr ➔ Descr
| sumD   : Π c: C. (I c ➔ Descr) ➔ Descr
| sigD   : Π n: Nat. (Fin n ➔ Descr) ➔ Descr

Decode : ★ ➔ Descr ➔ ★
Decode ·T idD           = T
Decode ·T constD        = Unit
Decode ·T (pairD d1 d2) = Pair ·(Decode ·T d1) ·(Decode ·T d2)
Decode ·T (sumD c f)    = Sigma ·(I c) ·(λ i: I c. Decode ·T (f i))
Decode ·T (sigD n f)    = Sigma ·(Fin n) ·(λ i: Fin n. Decode ·T (f i))

U : Descr ➔ ★
U d = μ (λ T: ★. Decode ·T d)

inSig : ∀ n: Nat. ∀ cs: Fin n ➔ Descr. Π i: Fin n. U (cs i) ➔ U (sigD n cs)
inSig -n -cs i d = in (i , d)
\end{verbatim}
  \caption{A closed universe of strictly positive types}
  \label{fig:closed-universe-def}
\end{figure}

The preceding section described an example of arity-generic programming.
We consider now a \emph{type-generic} task: proving the
\emph{no confusion} property~\cite{BG82_Algebras-Theories-Freeness} of datatype
constructors (that is, they are injective and disjoint) for a closed universe of
strictly positive types.
For defining a universe of datatypes, the idea (describe in more detail by
Dagand and McBride~\cite{DM12_Elab-Inductive-Definitions}) is to define a type
whose elements are interpreted as codes for datatype signatures and combine this
with a type-level least fixedpoint operator.

This universe is shown in Figure~\ref{fig:closed-universe-def}, where
\(\mathit{Descr}\) is the type of codes for signatures, \(\mathit{Decode}\) the
large elimination interpreting them, and \(C : \star\) and \(I : C \to \star\)
are parameters to the derivation.
Signatures comprise the identity functor (\(\mathit{idD}\)), a constant functor
returning the unitary type \(\mathit{Unit}\) (\(\mathit{constD}\)), a product of
signatures (\(\mathit{pairD}\)), and two forms of sums.
The latter of these, \(\mathit{sigD}\), is to be used to choose one of the
datatype constructors.
It takes an argument \(n : \mathit{Nat}\) (the number of datatype constructors)
and a family of \(n\) descriptions of the constructor argument types
(\(\mathit{Fin}\ n\) is the type of natural numbers less than \(n\)).
The former, \(\mathit{sumD}\), is a more generalized form that takes a code \(c
: C\) for a constructor argument type, and a mapping of values of type \(I\ c\)
(where \(I\) interprets these codes) to descriptions.
Both are interpreted by \(\mathit{Decode}\) as dependent pairs which pack
together an element of the indexing type (\(I\ c\) or \(\mathit{Fin}\ n\)) with
the decoding of the description associated with that index.

\begin{remark}
  \label{rem:data-constructor-type-args}
  In order to express a variety of datatypes, our universe is
  parameterized by codes \(C\) and interpretations \(I : C \to \star\) for
  constructor argument types, such as used in Example~\ref{ex:universe-list}.
  Unlike much of the literature describing the definition of a closed universe
  of strictly positive types~\cite{CDMM10_The-Gentle-Art-of-Levitation,
    DM12_Elab-Inductive-Definitions, Da13_Cosmology-of-Datatypes} wherein the
  host language is a variation of intrinsically typed Martin-L\"of type theory,
  CDLE is \emph{extrinsically typed} --- type arguments to constructors can play
  no role in computation, \emph{even} in the (simulated) computation of other types.
  This appears to be essential for avoiding paradoxes of the form described by
  Coquand and Paulin~\cite{CP88_Inductively-Defined-Types}, as CDLE is an
  impredicative theory in which datatype signatures need not be strictly
  positive.
\end{remark}

Finally, the family of datatypes within this universe is given as \(U\), defined
using a type-level least fixedpoint operator \(\mu\)
which we discuss in more detail in Section~\ref{sec:generic-large-elim}.
We define a constructor \(\mathit{inSig}\) for datatypes whose signatures are
described by codes of the form \(\mathit{sigD}\ n\ cs\) (for \(n : \mathit{Nat}\) and
\(\mathit{cs} : \mathit{Fin}\ n \to \mathit{Descr}\)) using the generic
constructor \(\mathit{in} : F\ \mu F \to \mu F\).

\begin{figure}
  \centering
  \small
\begin{verbatim}
data DecodeR (T: ★) : Descr ➔ ★ ➔ ★
= decIdR    : ∀ X: ★. TpEq ·X T ➾ DecodeR idD ·X
| decConstR : ∀ X: ★. TpEq ·X ·Unit ➾ DecodeR constD ·X
| decPairR  : ∀ d1: Descr. ∀ Ih1: ★. DecodeR d1 ·Ih1 ➔
              ∀ d2: Descr. ∀ Ih2: ★. DecodeR d2 ·Ih2 ➔
              ∀ X: ★. TpEq ·X ·(Pair ·Ih1 ·Ih2) ➾
              DecodeR (pairD d1 d2) ·X
| decSumR   : ∀ c: C. ∀ f: I c ➔ Descr. ∀ Ih: I c ➔ ★.
              (Π i: I c. DecodeR (f i) ·(Ih i)) ➔
              ∀ X: ★. TpEq ·X ·(Sigma ·(I c) ·Ih) ➾ DecodeR (sumD c f) ·X
| decSigR   : ∀ n: Nat. ∀ f: Fin n ➔ Descr. ∀ Ih: Fin n ➔ ★.
              (Π i : Fin n. DecodeR (f i) ·(Ih i)) ➔
              ∀ X: ★. TpEq ·X ·(Sigma ·(Fin n) ·Ih) ➾ DecodeR (sigD n f) ·X
\end{verbatim}
  \caption{Relational encoding of \(\mathit{Decode}\)}
  \label{fig:decoder}
\end{figure}

\begin{note*}
  For comparison with \(\mathit{Decode}\), the corresponding relational encoding
  used in our implementation is given in Figure~\ref{fig:decoder} as
  \(\mathit{DecodeR}\).
  Since the type argument \(T\) does not vary through recursive calls
  in \(\mathit{Decode}\), we have made it a parameter rather than index to
  \(\mathit{DecodeR}\) (in Cedille, recursive occurrences of the datatype being
  defined are not written applied to parameters; one writes e.g.
  \(\mathit{DecodeR}\ \mathit{id} \cdot X\) rather than \(\mathit{DecodeR} \cdot
  T\ \mathit{id} \cdot X\) in the type declarations of constructors).
  Aside from this, the method of translation from the syntax of pattern matching
  and recursion for large eliminations to GADTs with type equality constraints
  is the same as used in Section~\ref{sec:nary}.
\end{note*}

\begin{example}[Natural numbers]
  The type of natural numbers can be defined as:
\begin{verbatim}
unatSig : Descr
unatSig = sigD 2 (fvcons constD (fvcons idD fvnil))

UNat = U unatSig
\end{verbatim}
  \noindent where \(\mathit{fvcons}\) and \(\mathit{fvnil}\) are utilities for
  expressing functions out of \(\mathit{Fin}\ n\) in a list-like notation.
  The constructors of \(\mathit{UNat}\) are:
\begin{verbatim}
uzero : UNat
uzero = inSig fin0 unit

usuc : UNat ➔ UNat
usuc n = inSig fin1 n
\end{verbatim}
\end{example}

\begin{example}[Lists]
  \label{ex:universe-list}
  Let \(T : \star\) be an arbitrary type, and let parameters \(C\) and \(I\) be
  resp.\ \(\mathit{Unit}\) and \(\absu{\lambda}{\_}{T}\).
  The type of lists containing elements of type \(T\) is defined as:
\begin{verbatim}
ulistSig : Descr
ulistSig = sigD 2 (fvcons constD (fvcons (sumD unit (λ _. idD)) fvnil))

UList = U ulistSig
\end{verbatim}
  \noindent with constructors defined similarly to those of \(\mathit{UNat}\) in
  the preceding example.
\end{example}

\begin{figure}
  \centering
  \small
\begin{verbatim}
NoConfusion : Π n: Nat. Π cs: Fin n ➔ Descr. U (sigD n cs) ➔ U (sigD n cs) ➔ ★
NoConfusion n cs (in (i1 , d1)) (in (i2 , d2)) | i1 =? i2
NoConfusion n cs (in (i1 , d1)) (in (i1 , d2)) | yes eq = { d1 ≃ d2 }
NoConfusion n cs (in (i1 , d1)) (in (i2 , d2)) | no neq = False

inSigInj
: ∀ i1: Fin n. ∀ d1: Decode ·U (cs i1). ∀ i2: Fin n. ∀ d2: Decode ·U (cs i2).
  { inSig i1 d1 ≃ inSig i2 d2 } ➾ Pair ·{ i1 ≃ i2 } ·{ d1 ≃ d2 }

noConfusion : ∀ n: Nat. ∀ cs: Fin n ➔ Descr.
              Π d1: U (sigD n cs). Π d2: U (sigD n cs).
              { d1 ≃ d2 } ➔ NoConfusion d1 d2
noConfusion -n -cs (in (i1 , d1)) (in (i2 , d2)) eq | i1 =? i2
noConfusion -n -cs (in (i1 , d1)) (in (i2 , d2)) eq | yes eq' =
  snd (inSigInj -i1 -d1 -i2 -d2 eq)
noConfusion -n -cs (in (i1 , d1)) (in (i2 , d2)) eq | no neq' =
  neq' (fst (inSigInj -i1 -d1 -i2 -d2 eq))
\end{verbatim}
  \caption{Statement and proof of \emph{no confusion}}
  \label{fig:noconfusion}
\end{figure}

\paragraph*{Proving \emph{No Confusion}}
Figure~\ref{fig:noconfusion} shows the statement and
proof of the \emph{no confusion} property of datatype constructors.
\(\mathit{NoConfusion}\) is defined by case analysis over the two datatype
values, and additionally abstracts over a test of equality between \(\mathit{i1}\) and
\(\mathit{i2}\) to determine whether those values were both formed by the same
constructor.
The clause for which they are equal corresponds to the statement of constructor
injectivity (the two terms are equal only if equal arguments were given to the
constructor); the clause where \(\mathit{i1} \neq \mathit{i2}\) gives the
statement of disjointness (datatype expressions cannot be equal and also be in
the image of distinct constructors).
The proof \(\mathit{noConfusion}\) proceeds by abstracting over the same
equality test, using the lemma \(\mathit{inSigInj}\) that \(\mathit{inSig}\) is
injective (definition omitted; it follows from injectivity of both \(\mathit{in}\)
and the constructor for pairs) to finish the two cases. 

\subsection{Type Inhabitation of Simple Types}
\label{sec:decide-stlc}

\begin{figure}
  \centering
  \small
\begin{verbatim}
data SimpleTp : ★
= baseTp  : C ➔ SimpleTp
| arrowTp : SimpleTp ➔ SimpleTp ➔ SimpleTp

Simple : SimpleTp ➔ ★
Simple (baseTp c)    = I c
Simple (arrowTp a b) = Simple a ➔ Simple b

explode : ∀ X: ★. False ➔ X

Not : ★ ➔ ★
Not A = A ➔ False

decide : Π t: SimpleTp. (Π c: C. Sum ·(I c) ·(Not ·(I c))) ➔
         Sum ·(Simple t) ·(Not ·(Simple t))
decide (baseTp c) ctx = ctx c
decide (arrowTp a b) ctx | decide b ctx | decide a ctx
decide (arrowTp a b) ctx | in1 witB     | _        = in1 λ _. witB
decide (arrowTp a b) ctx | in2 notB     | in1 witA = in2 λ f. notB (f witA)
decide (arrowTp a b) ctx | in2 notB     | in2 notA = in1 λ x. explode (notA x)
\end{verbatim}
  \caption{Decision procedure for type inhabitation of a universe of simple types}
  \label{fig:stlc-codes}
\end{figure}

In this section we describe a decision procedure for type
inhabitation for a universe of simple types.
Figure~\ref{fig:stlc-codes} lists the datatype of codes
(\(\mathit{SimpleTp}\)) and its decoding by the large elimination
(\(\mathit{Simple}\)).
Like the previous section, the code is parametric in a type \(C\) of
codes for base types and an interpretation \(I : C \to \star\) mapping
them to Cedille types.

When inhabitation of the base types is decidable, to decide inhabitation of
a simple type it suffices to case split on the subdata. 
This decision procedure is \(\mathit{decide}\) in Figure~\ref{fig:stlc-codes}.
In the base case (\(\mathit{baseTp}\)) the decidability of the base type is
by assumption. 
The proof for the arrow case (\(\mathit{arrowTp}\)) is in the style of classical
logic, using the equivalence \(A \to B \equiv \neg A \lor B\).
Thus, we decide inhabitation of an arrow type by performing case analysis on
inhabitation of its domain and codomain types.

% Chris: I think that the below is wrong.
% For example, in the original code listing you had the type of ctx as
% > (∀ n: Nat. Sum ·(Ctx n) ·(Not ·(Ctx n)))
% So, in particular |ctx -0| = |ctx -1|.
% Let Ctx be such that Ctx 0 = ⊥ and Ctx 1 = ⊤
% Now perform case analysis on ctx -0 and ctxt -1
% 0. ctx -0 must be inj2 t for some t : Not ·⊥
% 1. ctx -1 must be inj1 tt for tt : ⊤
% and now we have |inj1 tt| = |inj2 t|
%
% Utilizing codes relevantly is unlikely to be the desired route with large eliminations as described herein.
% Indeed, writing a function that inducts over a code defeats the purpose of the constant time simulated computation rules.
% However, a decision procedure for types \emph{requires} a relevant representation and, moreover, if the resulting type is irrelevant the induction over the code is also irrelevant when the theorem is applied.
% Although most applications will desire the codes to be erased this example demonstrates that this is not always possible.

\subsection{Arity-generic Map Operation}
\label{sec:zipwith}

The last case study we consider is an arity-generic vector operation that
generalizes \(\mathit{map}\).
We summarize the goal (Weirich and
Casinghino~\cite{WC10_Generic-Programming-with-Dependent-Types} give a more
detailed explanation): define a function which, for all \(n\) and families of
types \((A_i)_{i \in \{1 \cdots n + 1\}}\), takes an \(n\)-ary function of type
\(A_1 \to \ldots \to A_n \to A_{n+1}\) and \(n\) vectors of type \(\mathit{Vec}
\cdot A_i\ m\) (for arbitrary \(m\)), and produces a result vector of type
\(\mathit{Vec} \cdot A_{n + 1}\ m\).
Note that when \(n = 1\), this is the usual map operation, and when \(n = 2\)
it is \(\mathit{zipWith}\) (when \(n = 0\), we have \(\mathit{repeat} :
\abs{\Pi}{m}{\mathit{Nat}}{A_1 \to Vec \cdot A_1\ m}\)).

\subsubsection{Vectors of Types}
\label{sec:tyvec}

\begin{figure}
  \centering
\small
\begin{verbatim}
𝒌TpVec (n : Nat) = Fin n ➔ ★

TVNil : 𝒌TpVec zero
TVNil _ = ∀ X: ★. X.

TVCons : Π n: Nat. Π H: ★. Π L: 𝒌TpVec n ➔ 𝒌TpVec (succ n)
TVCons n ·H ·L zeroFin = H
TVCons n ·H ·L (succFin i) = L i 

TVHead : Π n: Nat. 𝒌TpVec (succ n) ➔ ★
TVHead n ·L = L zeroFin

TVTail : Π n: Nat. 𝒌TpVec (succ n) ➔ 𝒌TpVec n
TVTail n ·L i = L (succFin i)

TVMap : Π F: ★ ➔ ★. Π n: Nat. 𝒌TpVec n ➔ 𝒌TpVec n
TVMap ·F n ·L i = F ·(L i)

TVFold : Π F: ★ ➔ ★ ➔ ★. Π n: Nat. 𝒌TyVec (succ n) ➔ ★
TVFold ·F zero ·L     = TVHead zero ·L
TVFold ·F (succ n) ·L = F ·(TVHead n ·L) ·(TVFold n ·(TVTail (succ n) ·L))
\end{verbatim}
  \caption{Vectors of types}
  \label{fig:tyvec}
\end{figure}

Our first task is to represent \(\mathit{Nat}\)-indexed families --- i.e.,
length-indexed vectors --- of types.
For reasons discussed in Remark~\ref{rem:data-constructor-type-args},
it is not possible to define vectors of types which support lookup as a Cedille
datatype.
Instead, we use simulated large eliminations to define them as lookup functions
directly.
This definition, along with some operations, is shown in Figure~\ref{fig:tyvec}.

The kind of length \(n\) vectors of types, \(\mathit{𝒌TpVec}\ n\), is defined as a
function from \(\mathit{Fin}\ n\) to \(\star\).
For the empty type vector \(\mathit{TVNil}\), it does not matter what type we
give for the right-hand side of the equation as \(\mathit{Fin}\ \mathit{zero}\)
is uninhabited.
For \(\mathit{TVCons}\), we use a (non-recursive) large elimination of the given
index, returning the head type \(H\) if it is zero and performing a lookup in
the tail vector \(L\) otherwise.
The destructors \(\mathit{TVHead}\) and \(\mathit{TVTail}\) and the mapping
function \(\mathit{TVMap}\) are defined as expected.
The fold operation, \(\mathit{TVFold}\), is given as a large elimination of the
\(\mathit{Nat}\) argument; in the successor case, the recursive call is made on
the tail of the given type vector \(L\).

Unlike the previous examples of large eliminations we have considered,
\(\mathit{TVFold}\) computes a type \emph{constructor} (of kind
\(\mathit{{\kappa}TyVec}\ (\mathit{succ}\ n) \to \star\)).
We therefore show its relational encoding as \(\mathit{TVFoldR}\) in
Figure~\ref{fig:tvfoldr}.
As the figure suggests, little of the method of simulation described in
Section~\ref{sec:nary} needs changing to accommodate this higher-kinded type
constructor.

\begin{figure}
  \centering
  \small
\begin{verbatim}
data TVFoldR (F: ★ ➔ ★ ➔ ★): Π n: Nat. 𝒌TyVec (succ n) ➔ ★ ➔ ★
= tvFoldRZ : ∀ L: 𝒌TyVec num1. ∀ X: ★. TpEq ·X ·(TVHead zero ·L) ➾
             TVFoldR zero ·L ·X
| tvFoldRS : ∀ n: Nat. ∀ L: 𝒌TyVec (add num2 n).
             ∀ Ih: ★. TVFoldR n ·(TVTail (succ n) ·L) ·Ih ➔
             ∀ X: ★. TpEq ·X ·(F ·(TVHead (succ n) ·L) ·Ih) ➾
             TVFoldR (succ n) ·L ·X
\end{verbatim}
  \caption{Relational encoding of \(\mathit{TVFold}\)}
  \label{fig:tvfoldr}
\end{figure}

\begin{note*}
  As noted in Remark~\ref{rem:tpeq-subst}, \(\mathit{TpEq}\) does not admit
  a general substitution principle.
  Concerning \(\mathit{TVFold}\), this means if \(F\) is
  not congruent with respect to type equality in its second argument, then for
  types of the form \(\mathit{TVFold} \cdot F\ (\mathit{succ}\ (\mathit{succ}\ n))
  \cdot L\) we can simulate only one computation step.
\end{note*}

\subsubsection{\(\mathit{ArrTp}\) and \(\mathit{nvecMap}\)}
\label{sec:arrty}

\begin{figure}
  \centering
  \small
\begin{verbatim}
ArrTp : Π n: Nat. 𝒌TpVec (succ n) ➔ ★
ArrTp = TVFold ·(λ X: ★. λ Y: ★. X ➔ Y)

ArrTpVec m n ·L = ArrTp n (TVMap ·(λ A: ★. Vec ·A m) (succ n) ·L)

vrepeat : ∀ A: ★. Π n: Nat. A ➔ Vec ·A n
vapp    : ∀ A: ★. ∀ B: ★. ∀ n: Nat. Vec ·(A ➔ B) n ➔ Vec ·A n ➔ Vec ·B n

nvecMap : Π m: Nat. Π n: Nat. ∀ L: 𝒌TpVec (succ n). ArrTp n ·L ➔ ArrTpVec m n ·L
nvecMap m n ·L f = go n ·L (vrepeat m f)
  where
  go : Π n: Nat. ∀ L: 𝒌TpVec (succ n) ➔ Vec ·(ArrTp n ·L) ➔ ArrTpVec m n ·L
  go zero     ·L fs = fs
  go (succ n) ·L fs = λ xs. go n ·(TVTail (succ n) ·L) (vapp -m fs xs)
\end{verbatim}
  \caption{Arity-generic map}
  \label{fig:nvecMap}
\end{figure}

We are now ready to define the arity-generic vector operation
\(\mathit{nvecMap}\), shown in Figure~\ref{fig:nvecMap}.
We begin with \(\mathit{ArrTp}\), the large elimination that computes the
type \(A_1 \to \cdots \to A_n \to A_{n + 1}\) as a fold over a vector of types
\(L = (A_i)_{i \in \{1 \cdots n+1\}}\).
The type \(\mathit{Vec} \cdot A_1\ m \to \ldots \to \mathit{Vec} \cdot A_n\ m
\to \mathit{Vec} \cdot A_{n+1}\ m\) is then constructed simply by composing
\(\mathit{ArrTp}\ n\) with a map over \(L\) taking each entry \(A_i\) to the
type \(\mathit{Vec} \cdot A_i\ m\), shown in \(\mathit{ArrTpVec}\).

For \(\mathit{nvecMap}\), we use \(\mathit{vrepeat}\) to create \(m\) replicas
of the given \(n\)-ary function argument \(f\), then invoke the helper function
\(\mathit{go}\) which is defined by recursion over \(n\).
In the zero case, \(\mathit{fs}\) has type \(\mathit{Vec} \cdot (\mathit{TVHead}\
\mathit{zero} \cdot L)\ m\), which is equal to the expected type (by the
computation laws for \(\mathit{ArrTp}\); we can prove that \(\mathit{Vec}\)
respects type equality).
In the successor case, \(\mathit{fs}\) is a vector of functions whose type is
equal to
\[\mathit{TVHead}\ (\mathit{succ}\ n) \cdot L \to \mathit{ArrTp}\ n\ \cdot
  (\mathit{TVTail}\ (\mathit{succ}\ n) \cdot L)\]
\noindent and the expected type is
\[\mathit{Vec} \cdot (\mathit{TVHead}\ (\mathit{succ}\ n) \cdot L)\ m \to
  \mathit{ArrTpVec}\ m\ n \cdot (\mathit{TVTail\ (\mathit{succ}\ n) \cdot L})\]
\noindent so we assume such a vector \(\mathit{xs}\), use \(\mathit{vapp}\) to
apply each function of \(\mathit{fs}\) point-wise to the elements of
\(\mathit{xs}\), then recurse to consume the remaining arguments.

\begin{remark}
  \label{rem:tyvec-eq}
  The high-level syntax we use to present \(\mathit{TVCons}\) and
  \(\mathit{TVFold}\) obscures the fact that additional lemmas are needed to
  effectively use the latter on type vectors built with the former.
  These lemmas are \(\mathit{tvFoldZC'}\) and \(\mathit{tvFoldSC'}\) in
  Figure~\ref{fig:tvfold-cong}, and they may be understood as providing
  alternative computation laws for \(\mathit{TVFold}\) when it is applied to
  type vectors of the form \(\mathit{TVCons}\ n \cdot X \cdot L\) and to type
  constructors \(F\) that respect extensional type equality in both arguments.
  \begin{figure}
    \centering
    \small
\begin{verbatim}
RespTpEq2 : Π F: ★ ➔ ★ ➔ ★. ★
RespTpEq2 ·F = ∀ A1: ★. ∀ A2: ★. TpEq ·A1 ·A2 ➾
               ∀ B1: ★. ∀ B2: ★. TpEq ·B1 ·B2 ➾
               TpEq ·(F ·A1 ·B1) ·(F ·A2 ·B2)

tvFoldZC' : ∀ F: ★ ➔ ★ ➔ ★. RespTpEq2 ·F ➾
            ∀ X: ★. TpEq ·(TVFold zero ·(TVCons zero ·X ·TVNil)) ·X

tvFoldSC' : ∀ F: ★ ➔ ★ ➔ ★. RespTpEq2 ·F ➾
            ∀ n: Nat. ∀ X: ★. ∀ L: 𝒌TyVec (succ n).
            TpEq ·(TVFold (succ n) ·(TVCons (succ n) ·X ·L)) ·(F ·X ·(TVFold n ·L))
\end{verbatim}
    \caption{Variant computation laws for \(\mathit{TVFold}\)}
    \label{fig:tvfold-cong}
  \end{figure}
\end{remark}

\section{Generic Simulation}
\label{sec:generic-large-elim}

We now generalize the approach outlined in Section~\ref{sec:nary} and derive
simulated large eliminations \emph{generically} (here meaning
\emph{parametrically}) for datatypes.
For this, we first review Mender-style iteration and the generic framework of
Firsov et al.~\cite{FBS18_Efficient-Mendler-Style-Lambda-Encodings} for inductive
Mendler-style lambda encodings of datatypes in Cedille.
Then, for an arbitrary positive datatype signature we define a notion of a
Mendler-style algebra at the level of types, overcoming a technical difficulty arising from
Cedille's truncated sort hierarchy, and show a sufficient condition
for type-level algebras to yield a simulated large elimination.

\subsection{Mendler-style Iteration and Encodings}
We briefly review the datatype iteration scheme \emph{\`a la} Mendler.
Originally proposed by Mendler~\cite{Men87_Recursive-Types-2O} as a method of
impredicatively encoding datatypes, Uustalu and
Vene have shown that it forms the basis of an alternative categorical semantics
of inductive datatypes~\cite{UV99_Mendler-Inductive-Types}, and the same have
advocated the Mendler style of coding recursion as more idiomatic than the
classical formulation of structured recursion
schemes~\cite{UV00_Coding-Recursion-Mendler}.

\begin{definition}[Mendler-style iteration]
  \label{def:mendl-style-iter}
  Let \(F : \star \to \star\) be a positive type scheme.
  The datatype with signature \(F\) is \(\mu F\) with constructor
  \(\mathit{in} : F \cdot \mu F \to \mu F\).
  The Mendler-style iteration scheme for \(\mu F\) is described by the typing
  and computation law given for \emph{fold} below:

  \[
    \begin{array}{cc}
      \infer{
      \Gamma \vdash \mathit{fold} \cdot T\ a : \mu F \to T
      }{
      \Gamma \vdash T : \star
      \quad \Gamma \vdash a : \abs{\forall}{R}{\star}{(R \to T) \to F \cdot R
      \to T}
      }
      &
        \mathit{fold} \cdot T\ a\ (\mathit{in}\ d) \rightsquigarrow a \cdot \mu
        F\ (\mathit{fold \cdot T\ a})\ d
    \end{array}
  \]
\end{definition}

In Definition~\ref{def:mendl-style-iter}, the type \(T\) (the \emph{carrier}) is
the type of results we wish to compute, and the term \(a\) (the \emph{action})
gives a single step of a recursive function, and we call the two of them
together a Mendler-style \(F\)-algebra.
We understand the type argument \(R\) of the action as a kind of subtype of the
datatype \(\mu F\) --- specifically, a subtype containing only predecessors on
which we are allowed to make recursive calls.
The first term argument of the action, a function of type \(R \to T\), is the
handle for making recursive calls; in the computation law, it is instantiated to
\(\mathit{fold} \cdot T\ a\).
Finally, the last argument is an ``\(F\)-collections'' of predecessors of the
type \(R\); in the computation law, it is instantiated to the collection of
predecessors \(d : F\ \mu F\) of the datatype value \(\mathit{in}\ d\).

\subparagraph*{Generic framework for Mendler-style datatypes}
\label{sec:firsov-background}

\begin{figure}[h]
  \centering
  \[
    \begin{array}{c}
      \begin{array}{lll}
        \mathit{Monotonic} \cdot F
        & =
        & \abs{\forall}{X}{\star}{\abs{\forall}{Y}{\star}{\mathit{Cast} \cdot X
          \cdot Y \Rightarrow \mathit{Cast} \cdot (F \cdot X) \cdot (F \cdot Y)}}
        \\
        \mathit{PrfAlg} \cdot F\ m \cdot P
        & =
        & \abs{\forall}{R}{\star}{
           \abs{\forall}{c}{\mathit{Cast} \cdot R \cdot \mu F}{
        \\  & &
        \quad (\abs{\Pi}{x}{R}{P\ (\mathit{cast}\ \mhyph c\ x)}) \to
                \abs{\Pi}{\mathit{xs}}{F \cdot R}{P\ (in\ \mhyph m\ \mhyph c\ \mathit{xs})}}}
      \end{array}
          \\ \\
    \begin{array}{cc}
      \infer{
       \Gamma \vdash \mu F : \star
      }{
       \Gamma \vdash F : \star \to \star
      }
      &
        \infer{
         \Gamma \vdash \mathit{in}\ \mhyph m :
           \abs{\forall}{R}{\star}{\mathit{Cast} \cdot R \cdot \mu F \Rightarrow F \cdot R \to \mu F}
        }{
         \Gamma \vdash F : \star \to \star
         \quad \Gamma \vdash m : \mathit{Monotonic} \cdot F
        }
    \end{array}
      \\ \\
      \infer{
       \Gamma \vdash \mathit{out}\ \mhyph m : \mu F \to F \cdot \mu F
      }{
      \Gamma \vdash F : \star \to \star
      \quad \Gamma \vdash m : \mathit{Monotonic} \cdot F
      }
     \\ \\ 
        \infer{
         \Gamma \vdash \mathit{ind}\ \mhyph m : 
         \abs{\forall}{P}{\mu F \to \star}{
          \mathit{PrfAlg} \cdot F\ m \cdot P
          \to \abs{\Pi}{x}{\mu F}{P\ x}}
        }{
         \Gamma \vdash F : \star \to \star
         \quad \Gamma \vdash m : \mathit{Monotonic} \cdot F
      }
      \\ \\
      \begin{array}{lll}
        |\mathit{ind}\ \mhyph m \cdot P\ a\ (\mathit{in}\ \mhyph m \cdot
        R\ \mhyph c\ \mathit{xs})|
        & =_{\beta\eta}
        & |a \cdot R\ \mhyph c\ (\absu{\lambda}{x}{\mathit{ind}\ \mhyph m \cdot
          P\ a\ (\mathit{cast}\ \mhyph c\ x)})\ \mathit{xs}|
        \\ |\mathit{out}\ \mhyph m\ (\mathit{in}\ \mhyph m \cdot R\ \mhyph c\ \mathit{xs})|
        & =_{\beta\eta}
        & |\mathit{xs}|
      \end{array}
    \end{array}
  \]
  \caption{Axiomatic summary of the generic framework of Firsov et
    al.~\cite{FBS18_Efficient-Mendler-Style-Lambda-Encodings}} 
  \label{fig:firsov-generic}
\end{figure}

Figure~\ref{fig:firsov-generic} gives an axiomatic summary of the generic
framework of Firsov et al.~\cite{FBS18_Efficient-Mendler-Style-Lambda-Encodings}
for deriving efficient Mendler-style lambda encodings of datatypes with
induction.
In all inference rules save the type formation rule of \(\mu\), the datatype
signature \(F\) is required to be \(\mathit{Monotonic}\) (that is, positive).
\begin{itemize}
\item \(\mathit{in}\) is the datatype constructor.
  For the developments in this section, we find the Mendler-style
  presentation given in the figure more convenient than the classical type of
  \(\mathit{in}\).
  
\item \(\mathit{out}\) is the datatype destructor, revealing the
  \(F\)-collection of predecessors used to construct the given value.
  
\item \(\mathit{PrfAlg}\) is a generalization of Mendler-style algebras to
  dependent types.
  Compared to the earlier discussion:
  \begin{itemize}
  \item the carrier is a predicate \(P : \mu F \to \star\) instead of a type;
    
  \item the informal intuition that \(R\) is a subtype of the datatype \(\mu F\)
    is made explicit by requiring a type inclusion of the former into the
    latter; and
    
  \item given a handle for invoking the inductive hypothesis on predecessors of
    type \(R\) and an \(F\)-collection of such predecessors, a \(P\)-proof
    \(F\)-algebra action must show that \(P\) holds for the value constructed from
    these predecessors using \(\mathit{in}\).
  \end{itemize}
  
\item \(\mathit{ind}\) gives the induction principle: to prove a property
  \(P\) for an arbitrary term of type \(\mu F\), it suffices to give a
  \(P\)-proof \(F\)-algebra.
  
\end{itemize}

\subsection{Mendler-style Type Algebras}
\label{sec:generic-algty}

Like other (well-founded) recursive definitions, a large elimination can be
expressed as a fold of an algebra.
In theories with a universe hierarchy, expressing this
algebra is no difficult task: the signature \(F\) can be universe polymorphic so
that its application to either a type or kind is well-formed.
This is \emph{not} the case for Cedille, however, which has a truncated
hierarchy of sorts and no sort polymorphism.
More specifically, there is no way to express a classical \(F\)-algebra on the
level of types, e.g., a kind \((F\ \star) \to \star\), as it is not possible to
define a function on the level of kinds (which \(F\) would need to be).

Thankfully, this difficulty \emph{disappears} when the type algebra is expressed
in the Mendler style!
This is because \(F\) does not need to be applied to the kind (\(\star\)) of
previously computed types, only to the universally quantified type \(R\).
Instead, types are computed from predecessors using an assumption of kind \(R
\to \star\).

\begin{figure}[h]
  \centering
  \small
\begin{verbatim}
𝒌AlgTy = Π R: ★. Cast ·R ·μF ➔ (R ➔ ★) ➔ F ·R ➔ ★ .

AlgTyResp : 𝒌AlgTy ➔ ★
= λ A: 𝒌AlgTy.
  ∀ R1: ★. ∀ R2: ★. ∀ c1: Cast ·R1 ·μF. ∀ c2: Cast ·R2 ·μF.
  ∀ Ih1: R1 ➔ ★. ∀ Ih2: R2 ➔ ★.
  (Π r1: R1. Π r2: R2. { r1 ≃ r2 } ➔ TpEq ·(Ih1 r1) ·(Ih2 r2)) ➔
  Π xs1: F ·R1. Π xs2: F ·R2. { xs1 ≃ xs2 } ➔
  TpEq ·(A ·R1 c1 ·Ih1 xs1) ·(A ·R2 c2 ·Ih2 xs2) .
\end{verbatim}
  \caption{Mendler-style type algebras}
  \label{fig:algty}
\end{figure}

Figure~\ref{fig:algty} shows the definition \(\mathit{{\kappa}AlgTy}\) of the
kind of Mendler-style type algebras with carrier \(\star\) (henceforth we will
refer to the actions of type algebras simply as \emph{algebra}).
Just as in the concrete derivation of Section~\ref{sec:nary}, we require that
type algebras must respect type equality.
This condition is codified in the figure as \(\mathit{AlgTyResp}\), which says:
\begin{itemize}
\item given two subtypes \(R_1\) and \(R_2\) of \(\mu F\) (which need \emph{not}
  be equal),
  
\item and two inductive hypotheses \(\mathit{Ih}_1\) and \(\mathit{Ih}_2\) for
  computing types from values of type \(R_1\) and \(R_2\), resp.,
  
\item that return equal types on equal terms, then
  
\item we have that the type algebra \(A\) returns equal types on equal
  \(F\)-collections of predecessors (where the types of predecessors are resp.\
  \(R_1\) and \(R_2\)).
\end{itemize}

\begin{example}
  \label{ex:naryagl}
  Let \(\mathit{F} \cdot R = 1 + R\) be the signature of natural numbers
  with \(\mathit{zeroF} : \abs{\forall}{R}{\star}{F \cdot R}\) and
  \(\mathit{succF} : \abs{\forall}{R}{\star}{R \to F \cdot R}\) the
  signature's injections.
  The Mendler-style type algebra for \(n\)-ary functions over type \(T\)
  is:
\begin{verbatim}
NaryAlg : 𝒌AlgTy
NaryAlg ·R c Ih (zeroF ·R) = T
NaryAlg ·R c Ih (succF ·R n) = T ➔ Ih n
\end{verbatim}

  Inspecting this definition, we see it indeed satisfies the above
  condition on type algebras, with the proof sketch as follows.
  Assuming \(\mathit{xs_1}\) and \(\mathit{xs_2}\) such that \(\{ \mathit{xs_1}
  \simeq \mathit{xs}_2\}\), we may proceed by considering the cases where both
  are formed by the same injection.
  In the \(\mathit{zeroF}\) case, the algebra returns \(T\), which is equal to itself;
  in the \(\mathit{succF}\) case, we have \(\{\mathit{succF}\ x_1 \simeq
  \mathit{succF}\ x_2\}\) for some \(x_1 : R_1\) and \(x_2 : R_2\), so by
  injectivity of \(\mathit{succF}\) we have \(\{ x_1 \simeq x_2\}\).
  We conclude by using our assumption to obtain that \(\mathit{TpEq} \cdot
  (\mathit{Ih}_1\ x_1) \cdot (\mathit{Ih}_2\ x_2)\) and a lemma that the arrow
  type constructor respects type equality.
\end{example}

\begin{remark}
  \label{rem:algtyresp-erasure-check}
  We again note that, in the definition of \(\mathit{AlgTyResp}\), the two
  assumed subtypes \(R_1\) and \(R_2\) need not be equal.
  As a consequence, in order to satisfy this condition the type produced by the
  algebra should \emph{not} depend on its type argument \(R\).
  A high-level surface language implementation for large eliminations in Cedille
  could require that the bound type variable \(R\) only occurs in type
  arguments of term subexpressions.
  As definitional equality of types is modulo erasure of typing annotations in
  term subexpressions, this would ensure that the meaning (extent) of the type
  does not depend on \(R\).
\end{remark}

\subsection{Relational Folds of Type Algebras}
\label{sec:generic-largeR}

\begin{figure}
  \centering
  \small
\begin{verbatim}
data FoldR : μF ➔ ★ ➔ ★
= foldRIn
  : ∀ R: ★. ∀ c: Cast ·R ·μF. ∀ xs: F ·R.
    ∀ Ih: R ➔ ★. (Π x: R. FoldR (cast -c x) ·(Ih x)) ➔
    ∀ X: ★. TpEq ·X ·(A ·R c ·Ih xs) ➾ FoldR (in -c xs) ·X

Fold : μF ➔ ★
Fold x = ∀ X: ★. FoldR x ·X ➾ X .

foldRResp   : ∀ x: μF. ∀ X1: ★. FoldR x ·X1 ➔ ∀ X2: ★. TpEq ·X1 ·X2 ➾ FoldR x ·X2
foldRUnique : ∀ x: μF. ∀ X1: ★. FoldR x ·X1 ➔ ∀ X2: ★. FoldR x ·X2 ➔ TpEq ·X1 ·X2
foldREx     : Π x: μF. FoldR x ·(Fold x)
\end{verbatim}
  \caption{Generic large elimination}
  \label{fig:generic-largeR}
\end{figure}

Figure~\ref{fig:generic-largeR} gives the definition of \(\mathit{FoldR}\), a
GADT expressing the fold of a type level algebra \(A : \mathit{𝒌AlgTy}\) over
\(\mu F\) as a functional relation (\(A\) and \(F\) are parameters to the
definition).
It has a single constructor, \(\mathit{foldRIn}\), corresponding to the single
generic constructor \(\mathit{in}\) of the datatype, whose type we read as
follows:
\begin{itemize}
\item given a subtype \(R\) of \(\mu F\) and a collection of predecessors
  \(\mathit{xs} : F \cdot R\), and
  
\item a function \(\mathit{Ih} : R \to \star\) that, for every element \(x\) in
  its domain, produces a type related (by \(\mathit{FoldR}\)) to that element, then
  
\item the datatype value constructed from \(\mathit{xs}\) is related to all
  types that are equal to \(A \cdot R\ c \cdot \mathit{Ih}\ \mathit{xs}\).
\end{itemize}

Just as in Section~\ref{sec:nary}, to show that the inductive relation given by
\(\mathit{FoldR}\) determines a function (from \(\mu F\) to equivalence classes
of types), we define a canonical name (\(\mathit{Fold}\)) for the types
determined by the datatype elements and prove that the relation satisfies three
properties: it respects type equality, and every datatype element (uniquely)
determines a type. 
The proofs of respectfulness and existence properties proceed similarly to the
concrete proofs given for \(n\)-ary functions (see the code repository for full
details).
In the proof of uniqueness, shown in Figure~\ref{fig:foldr-unqiue}, we use the
condition on type algebras.

\begin{figure}[h]
  \centering
  \small
\begin{verbatim}
foldRUnique : ∀ x: μ F. ∀ T1: ★. FoldR x ·T1 ➔ ∀ T2: ★. FoldR x ·T2 ➔ TpEq ·T1 ·T2
foldRUnique -([in ·R1 -c1 xs1 , in ·R2 -c2 xs2])
            ·T1 (foldR ·R1 -c1 xs1 ·Ih1 rec1 ·T1 -eqT1)
            ·T2 (foldR ·R2 -c2 xs2 ·Ih2 rec2 ·T2 -eqT2)
  = trans -eqT1 -(trans -eqA -(sym -eqT2))
  where
  ih : Π r1: R1. Π r2: R2. { r1 ≃ r2 } ➔ TpEq ·(X1 r1) ·(X2 r2)
  ih [r1 , r2] r2 β = foldRUnique -(cast -c1 r1) ·(Ih1 r1) (rec1 r1) ·(Ih2 r2) (rec2 r2)

  eqA : TpEq ·(A ·R1 c1 ·Ih1 xs1) ·(A ·R2 c2 ·Ih2 xs2)
  eqA = AC ·R1 ·R2 -c1 -c2 ·Ih1 ·Ih2 ih xs1 xs2 β
\end{verbatim}
  \caption{Uniqueness proof for \(\mathit{FoldR}\)}
  \label{fig:foldr-unqiue}
\end{figure}

\begin{proof}[Proof idea (uniqueness)]
  By induction on the two \(\mathit{FoldR}\) arguments, we know that the
  argument \(x : μ F\) has the form \(\mathit{in} \cdot R_1\ \mhyph c_1\
  \mathit{xs}_1\) and also the form \(\mathit{in} \cdot R_2\ \mhyph c_2\
  \mathit{xs}_2\) (this is what the \([ - , -]\) notation means), where
  \(\mathit{xs}_1 : F \cdot R_1\) and \(\mathit{xs}_2 : F \cdot R_2\).
  We therefore have \(|\mathit{xs}_1| =_{\beta\eta} |\mathit{xs}_2|\) (but not
  that \(R_1\) and \(R_2\) are equal).

  To make use of the assumption \(\mathit{AC} : \mathit{{\kappa}AlgTyResp} \cdot
  A\) (a parameter to the derivation), we must show that all equal terms of the
  two subtypes \(R_1\) and \(R_2\) are mapped by \(\mathit{Ih}_1\) and
  \(\mathit{Ih}_2\) to equal types.
  This is obtained by invoking the inductive hypothesis on values returned by
  each \(\mathit{rec}_i : \abs{\Pi}{x}{R_i}{\mathit{FoldR}\ (\mathit{cast}\
    \mhyph c_i) \cdot (\mathit{Ih}_i\ x)}\) (for \(1 \leq i \leq 2\)) revealed
  by pattern matching
  (in \(\mathit{ih}\), we again use the notation \([ r_1 , r_2]\) to indicate
  that pattern matching on the proof of equality gives us that \(r_1\) and
  \(r_2\) are equal, but \emph{not} that \(R_1\) and \(R_2\) are).
\end{proof}

\begin{remark}
  At present, we are unable to express in a \emph{single definition} folds over
  type-constructor algebras with arbitrarily kinded carriers.
  Thus, while this result is parametric in a datatype signature, it must be
  repeated once for each type constructor kind.
  This process is however entirely mechanical, so an implementation
  of a higher-level surface language for large eliminations in Cedille could
  elaborate each variant of the derivation as needed, removing the burden
  of writing boilerplate code.
\end{remark}

\section{Related Work}
\label{sec:related}

\subparagraph*{CDLE}
In an earlier formulation of CDLE~\cite{Stu17_CDLE}, Stump proposed a mechanism
called \emph{lifting} which allowed simply typed terms to be lifted to the level
of types.
While adequate for both proving constructor disjointness for natural numbers and
enabling some type-generic programming (such as formatted printing in the style of
\texttt{printf}), its presence significantly complicated the meta-theory of CDLE
and its expressive ability was found to be
incomplete~\cite{SJ20_Cedille-Syntax-Semantics}.
Lifting was subsequently removed from the theory, replaced with the simpler
\(\delta\) axiom for proof discrimination.

Marmaduke et al.~\cite{MJS20_Zero-Cost-Constructor-Subtyping} described a method
of encoding datatype signatures that enables constructor subtyping (\emph{\`a
  la} Barthe and Frade~\cite{BF99_Constructor-Subtyping}) with zero-cost type
coercions.
A key technique for this result was the use of intersection types and equational
constraints to simulate (again with type coercions) the computation of types by
case analysis on terms --- that is, non-recursive large eliminations.
Their method of simulation is therefore suitable for expressing type algebras,
but not their folds.

\subparagraph*{MLTT and CC}
Smith~\cite{Smi88_Independence-of-Peanos-4th-Axiom-from-MLTT} showed that
disjointness of datatype constructors was not provable in Martin-L\"of type
theory without large eliminations by exhibiting a model of types with only two
elements --- a singleton set and the empty set.
In the calculus of constructions,
Werner~\cite{Wer92_A-Normalization-Proof-for-an-Impredicative-Type-System-with-Large-Elims}
showed that disjointness of constructors would be contradictory by using an
erasure procedure to extract System~\(F^\omega\) terms and types, showing that a
proof of  \(1 \neq 0\) in CC would imply a proof of \((\abs{\forall}{X}{\star}{X \to
  X}) \to \abs{\forall}{X}{\star}{X}\) in \(F^\omega\).
\emph{Proof irrelevance} is central to both results.
Since in CDLE proof relevance is axiomatized with \(\delta\), this paper can be
viewed as a kind of converse to these results: large eliminations enable proof
discrimination, and proof discrimination together with extensional type equality
enable the simulation of large eliminations.

\subparagraph*{GADT Semantics}
Our simulation of large eliminations rests upon a semantics of GADTs which
(intuitively) interprets them as the least set generated by their constructors.
However, the semantics of GADTs is a subject which remains under investigation.
Johann and Polonsky~\cite{JP19+HO-Data-Types-Syntax-and-Semantics}
recently proposed a semantics which makes them functorial, but in which the
above-given intuition fails to hold.
In subsequent work, Johann et
al.~\cite{JGJ21_GADTs-Functoriality-Parametricity-Pick-Two} explain that GADTs
whose semantics are instead based on impredicative encodings (in which case they
are not in general functorial) may be equivalently expressed using explicit type
equalities.
Though they exclude functorial semantics for GADTs in CDLE, the presence of type
equalities (both implicit in the semantics and the explicit uses of derived
extensional type equality) are essential for defining a relational simulation of
large eliminations.

\section{Conclusion and Future Work}
\label{sec:conclusion}

We have shown that large eliminations may be simulated in CDLE using a derived
extensional type equality, zero-cost type coercions, and GADTs to inductively
define functional relations.
This result overcomes seemingly significant technical obstacles, chiefly CDLE's
lack of primitive inductive types and universe polymorphism, and is made possible
by an axiom for proof discrimination.
To demonstrate the effectiveness of the simulation, we examine several case
studies involving type- and arity-generic programming.
Additionally, we have shown that the simulation may be derived generically (that
is, parametric in a datatype signature) with Mendler-style type algebras
satisfying a certain condition with respect to type equality.

\subparagraph*{Syntax}
In this paper, we have chosen to present code examples using a high-level syntax
to improve readability.
While the current version of Cedille~\cite{Ced21_Cedille-Programming-Language}
supports surface language syntax for datatype declarations and recursion,
syntax for large eliminations remains future work.
Support for this requires addressing (at least) two issues.
First, it requires a sound criterion for determining when the
type algebra denoted by the surface syntax satisfies the condition
\(\mathit{AlgTyResp}\) (Section~\ref{sec:generic-algty}). 
We conjecture that a simple syntactic occurrence check, along the lines outlined
in Remark~\ref{rem:algtyresp-erasure-check}, for erased arguments will suffice.
Second, it is desirable that the type coercions that simulate the computation
laws of a large elimination be automatically inferred using a
subtyping system based on coercions~\cite{Lu99_Coercive-Subtyping,
  SHB09_A-Theory-of-Typed-Coercions}.

\subparagraph*{Semantics}
As discussed in Remark~\ref{rem:tpeq-subst}, the derived form of extensional
type equality used in our simulation lacks a substitution principle.
However, we claim that such a principle is validated by CDLE's
semantics~\cite{SJ20_Cedille-Syntax-Semantics}, wherein types are interpreted as
sets of (\(\beta\eta\)-equivalence classes of) terms of untyped lambda calculus.
Under this semantics, a proof of extensional type equality in the syntax implies
equality of the semantic objects.
We are therefore optimistic that CDLE may be soundly extended with a
kind-indexed family of type constructor equalities with an extensional
introduction form and substitution for its elimination form, removing all
limitations of the simulation of large eliminations.

%%
%% Bibliography
%%

%% Please use bibtex, 

\bibliography{biblio}

% \appendix

\end{document}